\documentclass{article}
\RequirePackage[a4paper]{geometry}
\usepackage{amsmath,graphicx}
\newcommand{\rr}{\mbox{\boldmath $r$}}
\newcommand{\EE}{\mbox{\boldmath $E$}}
\newcommand{\PPsi}{\mbox{\boldmath $\Psi$}}
\begin{document}
\renewcommand{\abstractname}{\vspace{-\baselineskip}}

\begin{center}
{\Large\bfseries
Using multidimensional speckle dynamics for high-speed, 
large-scale, parallel photonic computing
}\\

\vspace{3ex}
{\bfseries
Satoshi Sunada$^{1,2}$, Kazutaka Kanno$^3$, and Atsushi Uchida$^3$
\par}
{\footnotesize\itshape
$^1$Faculty of Mechanical Engineering, Institute of Science and
Engineering, Kanazawa University\\
Kakuma-machi Kanazawa, Ishikawa 920-1192, Japan \\
$^2$Japan Science and Technology Agency (JST), PRESTO, 4-1-8 Honcho,
 Kawaguchi, Saitama 332-0012, Japan
$^3$Department of Information and Computer Sciences, Saitama University,\\
255 Shimo-Okubo, Sakura-ku, Saitama City, Saitama, 338-8570, Japan.\\
\par}
\vspace{3ex}
\end{center}

\begin{abstract}
The recent rapid increase in demand for data processing has resulted in the
need for novel machine learning concepts and hardware. 
Physical reservoir computing and an extreme learning machine are novel
 computing paradigms based on physical systems themselves, where
the high dimensionality and nonlinearity play a crucial role in
 the information processing. 
Herein, we propose the use of multidimensional speckle dynamics in multimode
 fibers for information processing, where 
input information is mapped into the space, frequency, and time domains
by an optical phase modulation technique.
The speckle-based mapping of the input information is high-dimensional and
 nonlinear and can be realized at the speed of light; thus,  
nonlinear time-dependent information processing can successfully be achieved
at fast rates when applying a reservoir-computing-like-approach.  
As a proof-of-concept, we experimentally demonstrate chaotic time-series
 prediction at input rates of 12.5 Gigasamples per second. 
Moreover, we show that owing to the passivity of multimode fibers,
multiple tasks can be simultaneously processed within a single
 system, i.e., multitasking.  
These results offer a novel approach toward realizing parallel,
high-speed, and large-scale photonic computing.
\end{abstract}


\section{Introduction}
Reservoir computing (RC) \cite{Versraeten2007,Jaeger2004,Maass2002} and
an extreme learning machine (ELM) \cite{Huang2006} are
novel neuro-inspired computing methods that use high-dimensional 
and nonlinear systems, e.g., neural networks, to process input information.
Unlike a conventional neural network, internal networks are fixed, and
only the readout weights are trained, which greatly simplifies the training method 
and enables easy hardware implementation.
To date, a variety of physical implementations of RC and ELM, including
optoelectronic and photonic systems \cite{Ortin2015}, memristors \cite{Du2017}, 
spin waves \cite{Nakane2018}, and soft-robots \cite{Nakajima2015}, 
have been proposed, and have demonstrated a performance comparable with that of digital computing
based on other algorithms in a series of benchmark tasks, 
including time-series prediction, wireless channel equalization, and image and
phoneme recognition \cite{Tanaka2019}. 

Photonics-based RC is of particular interest. 
The use of photonic systems for RC implementation is expected to accelerate 
recurrent neural network processing with low energy costs \cite{GVderSande2017,Kitayama2019}. 
Most photonic RC systems are based on delayed feedback structures,
which can be simply realized using a nonlinear device with a time-delayed
feedback loop
\cite{Larger2017,Brunner2013,Larger2012,Uchida2020,Takano2018,Vinckier2015,Sugano2020}. 
Based on a time-multiplexing technique \cite{Appeltant2011,Paquot2012}, 
the sampled signals of the nonlinear device are used as virtual neurons
(network nodes) for information processing. 
With this approach, most complex tasks can be solved with few errors; however, 
these systems suffer from an inherent trade-off between the number of
neurons and the required processing time. 

Meanwhile, significant progress has been made with various
multiplexing approaches that utilize other degrees of freedom of light, i.e.,
space and wavelength, to overcome the above trade-off \cite{Vandoorne2014,Sunada2019,Laporte2018,Dong2020,Paudel2020,Bueno2018,Akrout2016}. 
Most multiplexing techniques are based on optical
data transportation techniques used for optical fiber communication 
and have the potential to realize high-speed parallel operation.   
Spatial multiplexing schemes, based on optical node arrays, have been
used for RC in silicon photonic chips and have demonstrated high-speed
processing of the bit header recognition at up to 12.5 Gbit/s
\cite{Vandoorne2014}. 
To generate a larger number of spatial nodes (i.e., a larger network
scale), 
the use of speckles generated
from light scattering \cite{Dong2020} and multimode waveguides
\cite{Paudel2020}, and diffractively coupled systems \cite{Bueno2018},
 have been proposed, demonstrating good performance 
even for complex tasks, such as chaotic time-series prediction. 
As other approaches, frequency multiplexing schemes have also
been proposed, where the amplitude and phase of the frequency sidebands of a
laser in a fiber loop are used as reservoir nodes, 
demonstrating nonlinear channel equalization and speech recognition
\cite{Akrout2016}. 
These approaches are promising; however, there are certain limitations.
For example, the constructions of the abovementioned spatial multiplexing networks
essentially involve a direct
trade-off between parallelism and footprint.  
In frequency multiplexing, the number of nodes may be practically
limited by the amplitudes or bandwidth of the electronic modulators.  

Herein, we propose a new approach to photonic parallel information
processing based on high-dimensional speckle dynamics generated from multimode fibers.
This approach easily enables the combined use of space, wavelength,
 and time multiplexing to achieve high-speed, large-scale, and parallel processing. 
In the proposed system, 
high nonlinearity required to process nonlinear tasks can be introduced 
using fast optical phase modulation over a few Gigasamples/s (GS/s). 
The origin of the nonlinearity is the nonlinear optical mapping 
from phase-encoded input information into speckle patterns.
This differs from other types of photonic RC using multimode fiber
   speckles \cite{Paudel2020}, where nonlinearity is electronically introduced.
As a proof-of-concept demonstration for high-speed parallel processing, 
we experimentally show that a chaotic time-series prediction
can be performed at 12.5 GS/s.
Moreover, the proposed approach enables
a multitasking operation, i.e., {\it simultaneous} processing of a
number of independent tasks with a {\it single} photonic processing unit, 
using both the space- and wavelength multiplexing
techniques together.
As a primitive demonstration, we also show the simultaneous processing of two
independent tasks with respect to nonlinear channel equalization.
 
\section{Speckle dynamics in multimode fibers}
\subsection{Multimode fiber speckles}
Multimode fibers support hundreds of guided modes with different phase
velocities. 
For monochromatic input light, 
a complex speckle pattern is generated in the intensity distribution at 
the end of the multimode fiber as a result of the interference of guided modes with 
different phase velocities \cite{Imai1980,Rawson1980}.
A remarkable feature of a speckle pattern is its dependence on the wavelength of
the input light. 
The longer the fiber length is, the more
sensitive it becomes \cite{Rawson1980,Redding2013}. 
This suggests that when the wavelength (frequency) of the
input light is modulated, the speckle pattern dynamically varies over
time. 
In other words, information encoded by the wavelength modulation of the input
light can be mapped as dynamically varying speckle patterns.
See the Appendix for the theoretical model.
 
\subsection{Experimental setup}
\begin{figure}[htbp]
\centering\includegraphics[width=12cm]{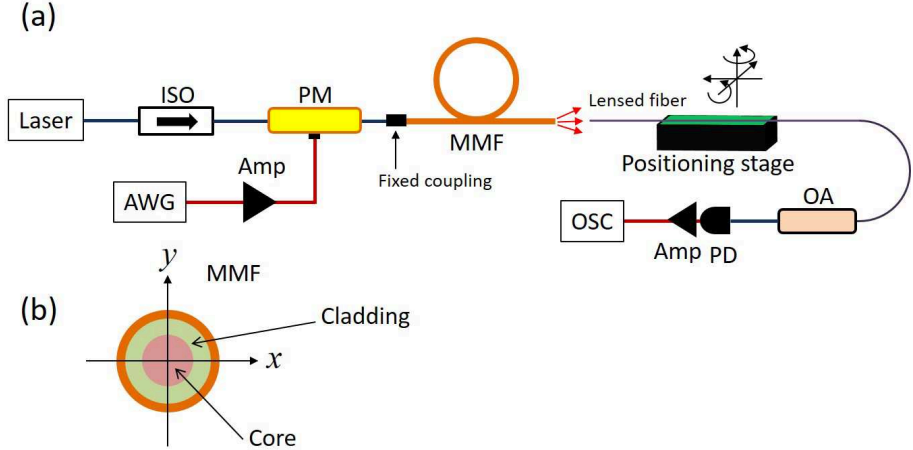}
\caption{\label{fig1}
(a) Experimental setup for generating speckle dynamics using a multimode
 fiber. 
ISO, optical isolator; Amp, electric amplifier; AWG, arbitrary waveform
 generator; PM, phase modulator;
 MMF, multimode fiber; OA, optical amplifier; PD, photodetector; OSC,
 digital oscilloscope.
A porlarization-maintaining single-mode fiber was coupled to the MMF through
 a standard FC/APC mating sleeve.
The fiber probe (lensed fiber) was adjusted to measure the speckle at the end of the
 multimode fiber using a five-axis positioning stage.
(b) Schematic of the cross-section at the end of the multimode fiber.
The origin of the $xy-$axis was set as the center of the core layer. 
}
\end{figure}

Figure \ref{fig1}(a) shows the experimental setup 
used to generate and measure the modulation-induced speckle dynamics in
a multimode fiber.
We used a commercially available step-index multimode fiber with a
core diameter of 50 $\mu$m, numerical aperture (NA) of 0.22, and length
of 10 m. 
The multimode fiber supports approximately 250 guided modes at an input
wavelength of $\lambda_0 =$ 1550 nm. 
The multimode fiber was covered with an insulator to reduce the effects
of thermal fluctuation and mechanical vibration.
A narrow-linewidth tunable laser (Alnair Labs, TLG-200, 100 kHz linewidth, 10 mW) 
was used as a light source. 
The laser light was phase-modulated using a phase modulator (EO Space,
AX-0MSS-20-LV, 16 GHz bandwidth) to encode an input signal, 
which was generated by an arbitrary waveform
generator (Tektronix, AWG70002A, 25 GS/s), as an instantaneous
frequency.
The maximum phase modulation was approximately 0.9$\pi$.
The modulated light was sent through a
polarization-maintaining single-mode fiber to the multimode fiber. 
We used a lensed fiber (spot size of $\sim 2$ $\mu$m with a working
distance of $\sim 10$ $\mu$m) as a probe to measure the time-variation of the output near the focal
position at the end of the multimode fiber, and the output was sent
to a photodetector (New Port, 1554-B, 12 GHz bandwidth) through an optical
fiber amplifier (FiberLabs, AMP-FL80133-CB) and measured using a digital
oscilloscope (Tektronix, DPO71604B, 50 GS/s, 16 GHz bandwidth). 

To obtain the time-variation of the speckle pattern,  
the measurement position was repeatedly changed by changing the position of
the lensed fiber in parallel with the cross-section of the fiber end
[Fig. \ref{fig1}(b)] with a micro-positioning stage, whereas
the input light was repeatedly modulated using the same signal.
The origins of the time variation of the output signals at each
measurement position were matched by adjusting each trigger signal
to reconstruct the intensity patterns at each time.
This method is effective because a multimode fiber is passive with
respect to the input,
and the output demonstrates consistency for the same inputs \cite{Uchida2004}.

\subsection{Speckle dynamics}
\begin{figure}[htbp]
\centering\includegraphics[width=12cm]{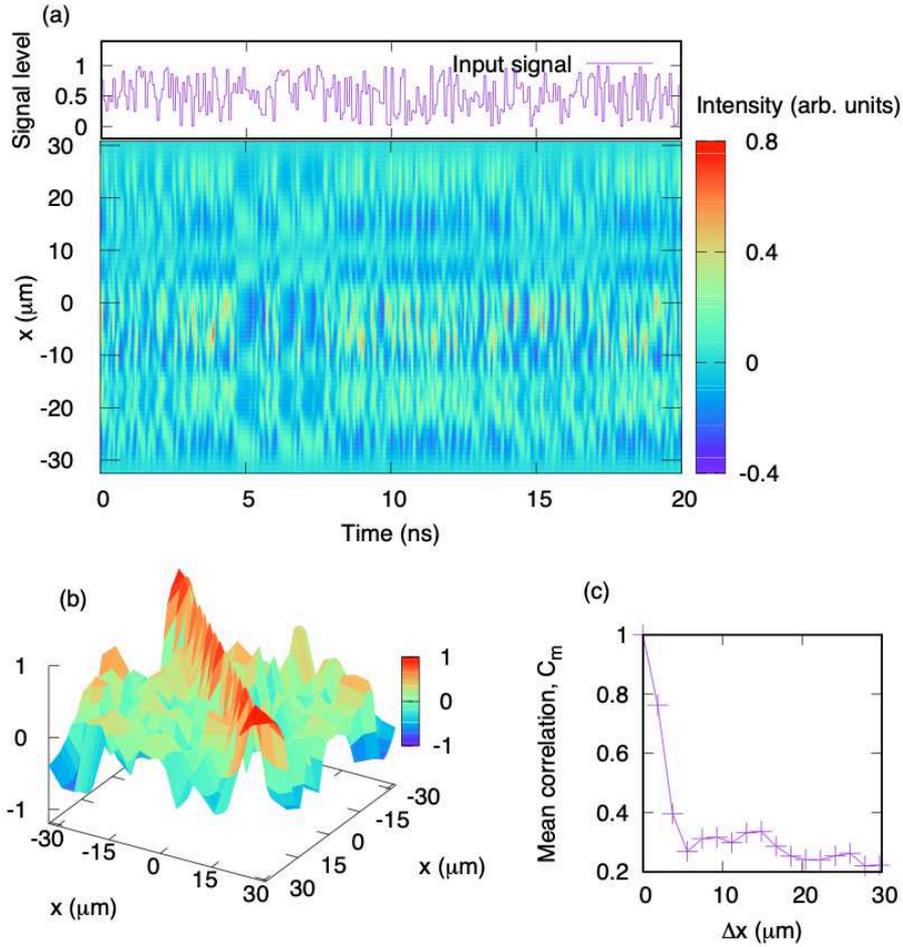}
\caption{\label{fig2}
(a) Input signal $u$ (upper panel) and output intensity
 signals $\phi^{\lambda_0}(x,t)$ (bottom panel)
 measured at the end of the multimode fiber by scanning using a lensed
 fiber probe along the $x-$axis shown in Fig. \ref{fig1}(b) 
(see Visualization 1).
The input rate $1/\tau$ was set to 12.5 GS/s.
The input signal $u$ is displayed as a function of time $t = n\tau$, and the origin of
 time ($t = 0$) is adjusted to that for the output signals for display purposes. 
(b) Correlation map $C(x,x')$ between the output signals
 $\phi^{\lambda_0}(x,t)$ and $\phi^{\lambda_0}(x',t)$, which were
 measured at fiber positions $x$ and $x'$, respectively. 
(c) Mean correlation $C_m$ as a function of $\Delta x = |x-x'|$. 
The correlation sufficiently decays when $\Delta x > 4$ $\mu$m.  
}
\end{figure}

Figure \ref{fig2}(a) shows an instance of the time variation of the
intensity pattern measured by the above method, where the phase of the
input light was modulated at 12.5 GS/s with a pseudorandom
sequence
and each intensity signal was measured along the $x-$axis shown in
Fig. \ref{fig1}(b).
The intensity pattern varies in a nonlinear manner, according to
the input signal. 
The nonlinearity originates from the relationship between the input signal
and output intensity signal, i.e., the input signal is encoded with 
the phase modulation, whereas the output is obtained as an intensity
signal (see the Appendix). 
Considering that the computational capacity of the system depends on 
the number of linearly independent signals measured in the system
\cite{Dambre2012}, 
we measured the
correlation between two signals measured at different positions, i.e.,
\begin{equation}
C(x,x')=\dfrac{
\langle [\phi^{\lambda_0}(x,t)-\bar{\phi}^{\lambda_0}_x]
[\phi^{\lambda_0}(x',t)-\bar{\phi}^{\lambda_0}_{x'}]
\rangle_T
}
{
\sigma_x\sigma_x'},
\end{equation} 
where $\phi^{\lambda_0}(x,t)$ represents the intensity measured at
position $x$ at time $t$ for input wavelength $\lambda_0$.
Here, $\langle f \rangle_T = 1/T \int^T_0 f dt$ denotes 
the mean of $f$ over measurement time $T$.
In addition, $\bar{\phi}_x^{\lambda_0}$ and $\sigma_x$ are the mean and standard
deviation of $\phi^{\lambda_0}(x,t)$, respectively.   
The correlation $C(x,x')$ decays for large $\Delta x =|x'-x|$, as
shown in Fig. \ref{fig2}(b).  
To clarify this, we calculated the mean correlation value
$C_m(\Delta x) = 1/D\int |C(x,x+\Delta x)| dx$ and
plot $C_m(\Delta x)$ in Fig. \ref{fig2}(c). 
Here, $C_m(\Delta x)$ sufficiently decays when $\Delta x > 4$ $\mu$m for the multimode 
fiber used in our experiment. 
We use the output signals sampled at the interval 
$\Delta x \approx 4$ $\mu$m for information processing, as discussed in
Sec. \ref{sec_method}. 
\subsection{Frequency dependence \label{sec_f}}
\begin{figure}[htbp]
\centering\includegraphics[width=12cm]{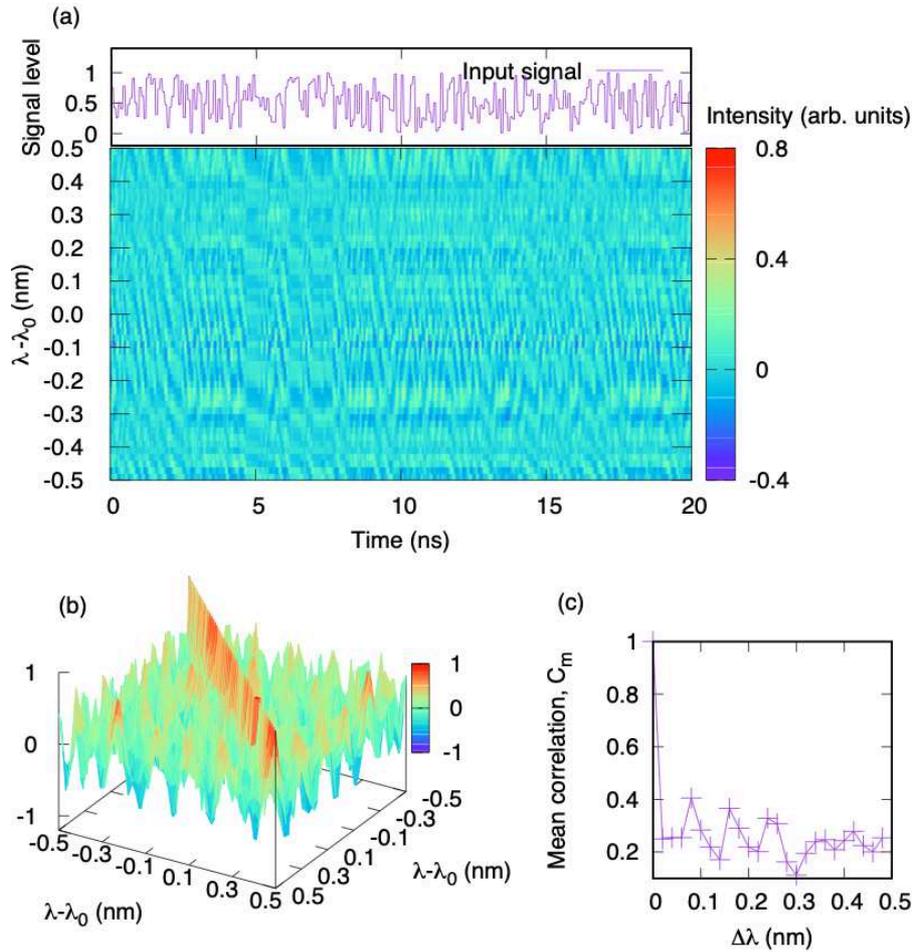}
\caption{\label{fig3}
(a) Time series of input signal $u$ (upper panel) and output intensity
 signals $\phi^{\lambda}(x_0,t)$ (bottom panel)
 measured at the fixed position $x_0$ by changing the wavelength
 $\lambda$ of the light source. 
In (a), the input signal $u$ is displayed as a function of time $t=n\tau$, and the origin of
 time ($t = 0$) is adjusted to that for the output signals for display purposes. 
(b) Correlation map $C(\lambda,\lambda^{'})$ between the output signals
 $\phi^{\lambda}(x_0,t)$ and $\phi^{\lambda'}(x_0,t)$.
(c) Mean correlation $C_m$ as a function of $\Delta \lambda = |\lambda-\lambda^{'}|$. 
The correlation sufficiently decays when $\Delta \lambda > 0.02$ nm.  
}
\end{figure}

The speckle dynamics depends on the wavelength (frequency) of the input light, 
and a variety of dynamics can be generated with different wavelengths in parallel.  
To confirm this, we varied the wavelength $\lambda$ of the input light
around $\lambda_0 = $ 1550 nm and measured the intensity variation from the
multimode fiber. 
In the results shown in Fig. \ref{fig3}(a), the input light is 
phase-modulated with the same pseudorandom sequence at 12.5 GS/s. 
As can be seen in the figure, various time variations are obtained, depending on
the input wavelength $\lambda$. 
We also measured the correlation, 
\begin{equation}
C(\lambda,\lambda^{'})=
\dfrac{
\langle 
\left[
\phi^{\lambda}(x_0,t)
-\langle \phi_{x_0}^{\lambda}\rangle 
\right]
[
\phi^{\lambda^{'}}(x_0,t)
-\langle \phi_{x_0}^{\lambda^{'}}\rangle 
]
\rangle_T
}
{
\sigma_\lambda\sigma_{\lambda^{'}}
}, 
\end{equation}
where 
$\phi^\lambda(x_0,t)$ represents the intensity measured at $x=x_0$ and
$y\approx 0$ for input wavelength $\lambda$ [Fig. \ref{fig3}(b)].
Figure \ref{fig3}(c) shows the mean correlation 
$C_m(\Delta\lambda)
 = 1/D_\lambda\int |C(\lambda,\lambda+\Delta\lambda)|d\lambda
$. 
Here, $C_m(\Delta\lambda)$ sufficiently decays when $\Delta\lambda >$ 0.02 nm
for the multimode fiber.
$\Delta\lambda$ of 0.02 nm corresponds to lower bounds for 
the parallel generation of different speckles and modulation 
bandwidth for information encoding with input lights of different wavelengths.
%
The correlation $C_m(\Delta\lambda)$ can more rapidly decay with an increase in the length of 
the multimode fiber \cite{Redding2013}, 
and a variety of dynamics that are more
sensitive to the wavelength can be obtained.
This result suggests that with multiple light sources of
different wavelengths, the input information can be encoded even within the
wavelength domain with a multiplexing technique, and parallel information
processing is enabled, as demonstrated in Sec. \ref{sec5}. 

\section{Information processing based on speckle dynamics}
\subsection{Method \label{sec_method}}
As demonstrated in the previous section, the use of a multimode fiber
enables high-dimensional and nonlinear mapping of the input
information in the speckle patterns.
We use the mapping properties based on multimode fibers to process time-dependent
input information in the following ways:
The input light is phase-modulated with a signal, $u(n)$
$(n \in \{1,2,\cdots\})$, where $u(n)$ holds for the time interval
$\tau$. 
The output intensity signals, $\phi_{i}^{\lambda_0}(t) = |E(\rr_i, t)|^2$, from
the multimode fiber are measured at $\rr_i = (x_i,y_i)$ with steps 
$\Delta x = \Delta y$ $\approx 4$ $\mu$m in the $xy$-plane [Fig.\ref{fig1}(b)] 
at the sampling time $t = n\tau$,
 where $i \in \{1, 2, \cdots, N \}$ is the index of the measurement
position. 
The readout is given by the following:
\begin{equation}
\hat{y}(n) =\sum_{i=1}^N w_i \phi^{\lambda_0}_i(n\tau), \label{eq_readout1}
\end{equation}
where $w_i$ is a readout weight. 
The goal of the processing is to approximate a
functional relationship between the input signal $u(n)$ and target $y(n)$
using $\hat{y}(n)$. 
To this end, a finite set of training data 
$\{u(n),y(n) \}_{n=0}^{T_n}$ is utilized to calculate the optimal readout weights. 
This is performed through a simple ridge regression to minimize the error,
$\sum_n^{T_n}|y(n)-\hat{y}(n)|^2 + \lambda\sum_i^N w_i^2$, 
where $\lambda$ is the regularization parameter to avoid an
ill-conditioned problem \cite{Larger2017}.

\subsection{Chaotic time-series prediction}
\begin{figure}[htbp]
\centering\includegraphics[width=15cm]{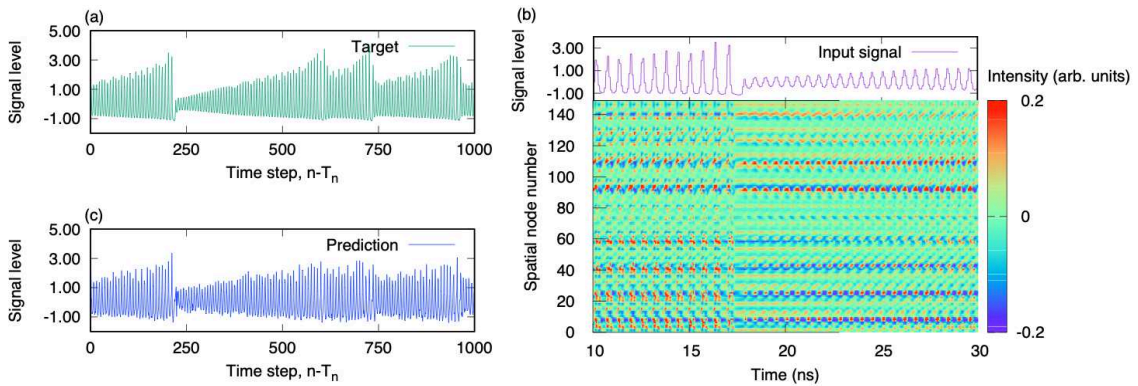}
\caption{\label{fig4}
(a) Target signal $y(n)$ in the Santa Fe time-series prediction task. 
(b) Input signal $u$ as a function of time $t = (n-T_n)\tau$ (upper panel), 
where $T_n=3000$ and $\tau = 0.08$ ns (input rate of $1/\tau=12.5$ GS/s).  
The lower panel shows the measured intensity signals,
 $[\phi_1^{\lambda_0}(t),\phi_2^{\lambda_0}(t),\cdots,\phi_N^{\lambda_0}(t)]$,
responding to the input light with wavelength $\lambda_0=1550$ nm, which
 is phase-modulated by the input signal $u$ (see Visualization
 2). These signals are used as nodes to obtain the readout $\hat{y}$
 [see Eq. (\ref{eq_readout1})].  
(c) The readout $\hat{y}(n)$ obtained using the measured 
signals, $[\phi_1^{\lambda_0}(t),\phi_2^{\lambda_0}(t),\cdots,\phi_N^{\lambda_0}(t)]$.
}
\end{figure}

 As a demonstration of a computation application using speckle dynamics, 
we conducted the Santa Fe time-series prediction task \cite{Weigend1993}, 
which is a one-step ahead prediction task of chaotic data generated from 
a far-infrared laser [see Fig. \ref{fig4}(a)]. 
In this task, the input signal $u(n)$ corresponds to the $n$-th sampling
point of the chaotic waveform, and $y(n)$ is set as the $n+1$-th sampling point, $u(n+1)$. 
 The prediction error was evaluated using the normalized mean square
 error (NMSE), which is given by
 $1/T_n\sum_{n=1}^{T_n}|y(n)-\hat{y}(n)|^2/\sigma_y^2$, where
 $\sigma_y^2$ is the variance of the target signal $y(n)$. 
We used $T_n =$ 3000 steps for training and 1000 steps for testing. 

Figures \ref{fig4}(b) and \ref{fig4}(c) show the results of the measured intensity signals $\phi_i^{\lambda_0}(t)$ 
and readout $\hat{y}(n)$ for an input rate of  
$1/\tau =$ 12.5 GS/s of the signal $u(n)$, respectively. 
According to Eq. (\ref{eq_readout1}), the readout $\hat{y}(n)$ was
calculated with $N =$ 150 intensity signals, which were 
measured at 60$\times$60 $\mu$m$^2$ near the fiber end.
The NMSE of the prediction was approximately 0.09, which is comparable with
the NMSEs obtained for the other photonic systems. (As an example, NMSEs of 0.106 and 0.109 have been reported for $N$ = 388 and 124 in \cite{Brunner2013} and \cite{Takano2018}, respectively.) 
Unlike in other photonic information processing systems 
using laser devices \cite{Brunner2013,Takano2018},  
active devices are not involved in the processing in the proposed system; thus, 
the response time is not limited.
We expect that faster speckle dynamics and then faster processing can be realized 
by increasing the input rate $1/\tau$.

\subsection{Using wavelength- and time-multiplexing techniques}
The number of nodes $N$ required to construct a readout $\hat{y}$ [Eq. (\ref{eq_readout1})] 
is limited by the physical dimensions of the multimode fiber. 
However, as shown in Fig. \ref{fig3}(a), 
the signals generated for different input wavelengths are expected to be used 
as additional nodes in the proposed system, 
enabling processing on a larger-scale network.
In addition, although the present experimental system does not have any memory
effect, which is crucial for time-dependent signal processing, the
memory effect can be introduced by adding an optical cavity in the
system \cite{Sunada2019} or using the past information as a node in the time
domain with a time-multiplexing method \cite{Appeltant2011}. 
In this study, we simply use the latter method, i.e, the time-multiplexing
method, and explore the availability of additional nodes in the time and
wavelength domains to discuss the potential improvement of the
computational performance. 
We here consider the following three readouts:
\begin{eqnarray}
\hat{y}_1(n)=\sum_{k=0}^K 
\sum_{i=1}^Nw_{i,k}\phi^{\lambda_0}_{i}[(n-k)\tau], \\
\hat{y}_2(n)=\sum_{k=0}^K
\sum_{j=1}^Mw_{j,k}\phi^{\lambda_j}_0[(n-k)\tau], \\
\hat{y}_3(n)=\sum_{k=0}^K
\left[
\sum_{i=1}^Nw_{i,k}\phi^{\lambda_0}_{i}[(n-k)\tau]+\sum_{j=1}^Mw_{j,k}\phi^{\lambda_j}_0[(n-k)\tau] 
\right], 
\end{eqnarray}
where
the output intensity $\phi_{i}^{\lambda_0}[(n-k)\tau]$ is 
sampled at a measurement position labeled by $i$ for the fixed input wavelength $\lambda_0$
at time $t = (n-k)\tau$.
Here, $\phi_{0}^{\lambda_j}(t)$ is the output intensity measured at $i =
0$ for input wavelength $\lambda_j$ ($j \in \{1,2,\cdots M\}$) at time $t$. 
In addition, $\hat{y}_1(n)$, $\hat{y}_2(n)$, and $\hat{y}_3(n)$ 
are the readouts in the space, wavelength, and mixed (space/wavelength)
domains, respectively, and 
$K$ denotes the number of past output signals used for
calculating the readouts. 
The readout weights $w_{i,k}$ and $w_{j,k}$ were trained for the Santa Fe time-series prediction task. 

\begin{figure}[htbp]
\centering\includegraphics[width=10cm]{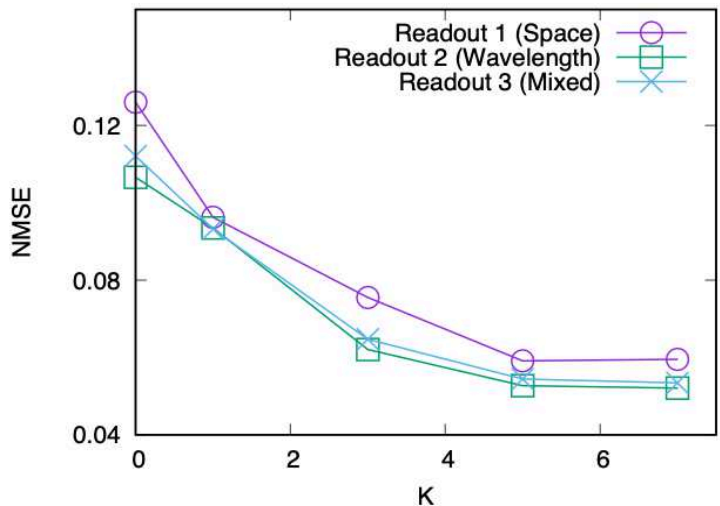}
\caption{\label{fig5}
Prediction errors (NMSEs) as a function of $K$ for three readouts,
$\hat{y}_1(n)$, $\hat{y}_2(n)$, and
 $\hat{y}_3(n)$, which 
are plotted using open circles, open squares, and crosses, respectively. 
The numbers of nodes used for readouts 1, 2, and 3
are $N = 50$, $M = 50$, and $N = M = 25$, respectively. 
}
\end{figure}

Figure \ref{fig5} shows the result of the prediction task, which was
obtained by the trained readouts $\hat{y}_l(n)$ ($l\in \{1,2,3\}$), 
where $\lambda_0 = 1550$ nm and $\lambda_j = 1549.5 + 0.02j$ nm. 
For comparison, the total number of nodes, $N+M$, is fixed at 50
for all cases.   
As $K$ increases, the NMSEs decrease and the prediction performance improves.   
The readout $\hat{y}_1(n)$ shows
the best NMSE of $0.06$ for $K = 7$, which is better than 
that of other photonic RC \cite{Brunner2013,Takano2018}.   
The readouts $\hat{y}_2(n)$ and $\hat{y}_3(n)$ exhibit a relatively improved
performance, and the best NMSE was 0.052.
This improvement may be attributed to the linear independence
among the signals $\phi^{\lambda_j}_0$ in the wavelength domain
[Fig. \ref{fig3}(a)]. 
These results suggest that 
the simultaneous use of the nodes in the space, wavelength, and time domains 
is effective for processing time-dependent signals. 
\section{Multitasking \label{sec5}}
\begin{figure}[htbp]
\centering\includegraphics[width=12cm]{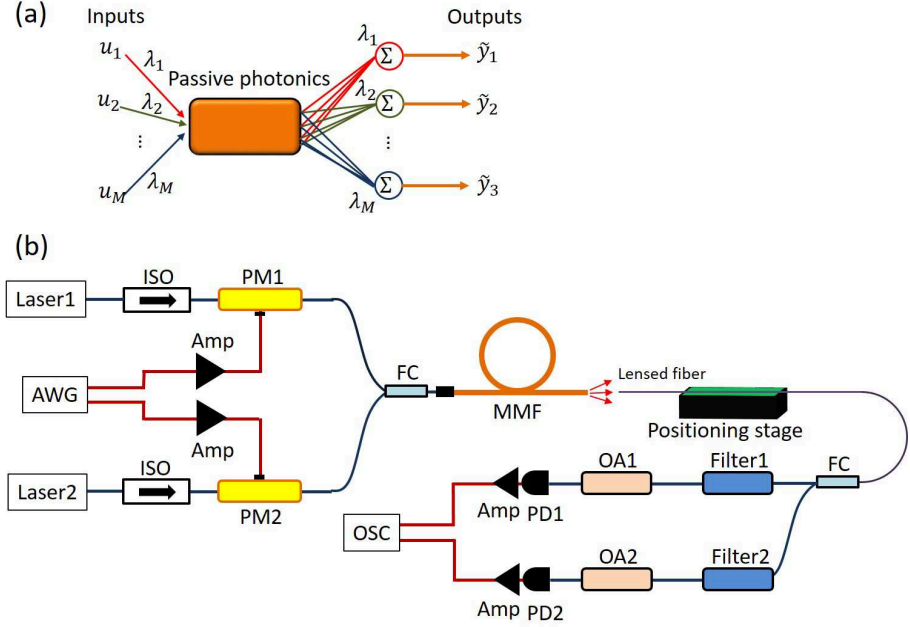}
\caption{\label{fig6}
(a) Schematic of multitasking based on passive photonics. 
Each input signal $u_j$  $(j \in \{1,2,\cdots, M\})$ 
is encoded as optical signals with wavelength $\lambda_j$
and is independently processed 
in a single photonic component (multimode fiber)
to obtain the readout $\tilde{y}_i$.   
(b) Experimental setup used for multitasking operation based on a
 multimode fiber.
Two input signals are generated from an arbitrary waveform generator
 (AWG).
The input lights with wavelengths $\lambda_1$ and
 $\lambda_2$ are phase-modulated using PM1 and PM2, respectively, 
and injected into a multimode fiber (MMF).  
The outputs from the MMF are separated by two optical filters (Filter 1
 and 2) and sent to photodetectors PD1 and PD2 through optical amplifiers
 (OA1 and OA2), respectively.   
}
\end{figure}

Finally, we discuss the simultaneous use of the above space and wavelength
multiplexing techniques, which also allows a multitasking operation, i.e., 
 parallel processing of numerous {\it independent} tasks, 
with a {\it single} photonic
component, that is, a multimode fiber [Fig. \ref{fig6}(a)].
This is enabled by the encoding of different input information 
at different wavelengths because there are no nonlinear interactions
between the lights with different wavelengths.
Figure \ref{fig6}(b) shows the experimental setup for the multitasking operation. 
We used two light sources (wavelength-tunable lasers), the wavelengths of which are $\lambda_1$ and
$\lambda_2$, respectively. 
The respective lights were then independently phase-modulated with
different input signals, $u_1$ and $u_2$, and injected into 
a multimode fiber. 
The output light of $\lambda_{1(2)}$ from the multimode fiber was detected by
PD${1(2)}$ after passing through a tunable optical filter (Filter ${1(2)}$) with
a bandwidth (full width at half maximum) of $\Delta \approx 1.1$ nm,
the center transmission wavelength of which was tuned at $\lambda_{1(2)}$ 
and used as the spatial nodes for calculating the readouts. 

As a demonstration, we used the nonlinear channel equalization task \cite{Jaeger2004}.
A goal of this task is to reconstruct four digital signals $\{-3,-1,1,3\}$ 
transmitted through a communication channel with nonlinear distortion. 
The nonlinear transformation of the communication channel is given by 
a model equation \cite{Paquot2012} 
\begin{eqnarray}
q(n) = && 0.08d(n+2)-0.12d(n+1)+d(n)+0.18d(n-1) \nonumber \\
&& -0.1d(n-2)+0.091d(n-3)-0.05d(n-4)  \\
&& +0.04d(n-5)+0.03d(n-6)+0.01d(n-7), \nonumber \\
u(n) = && q(n) + 0.036q(n)^2 -0.011q(n)^3 + v(n),
\end{eqnarray}
where $d(n) \in \{-3,-1,+1,+3\}$ is the original signal before
transmission through the communication channel, $q(n)$ is the linear
channel output, $u(n)$ is the noisy nonlinear channel output, 
and $v(n)$ is white Gaussian noise with a zero mean. 

In our experiment, we assumed 
two original signals, $d_1(n)$ and $d_2(n)$, consisting of 
two independent random sequences, and two different noisy communication
channels with different noises, $v_1(n)$ and $v_2(n)$.
The power ratio between the signal $d_i(n)$ ($i=1,2$) and noise $v_i(n)$ ($i=1,2$) 
was set to 30 dB.
The goal of our experiment is to {\it simultaneously} recover the two original
signals, $d_1(n)$ and $d_2(n)$, from the respective noisy nonlinear
channel outputs 
$u_1(n)$ and $u_2(n)$. 
The computation performance was evaluated using the symbol error rate (SER). 

\begin{figure}[htbp]
\centering\includegraphics[width=12cm]{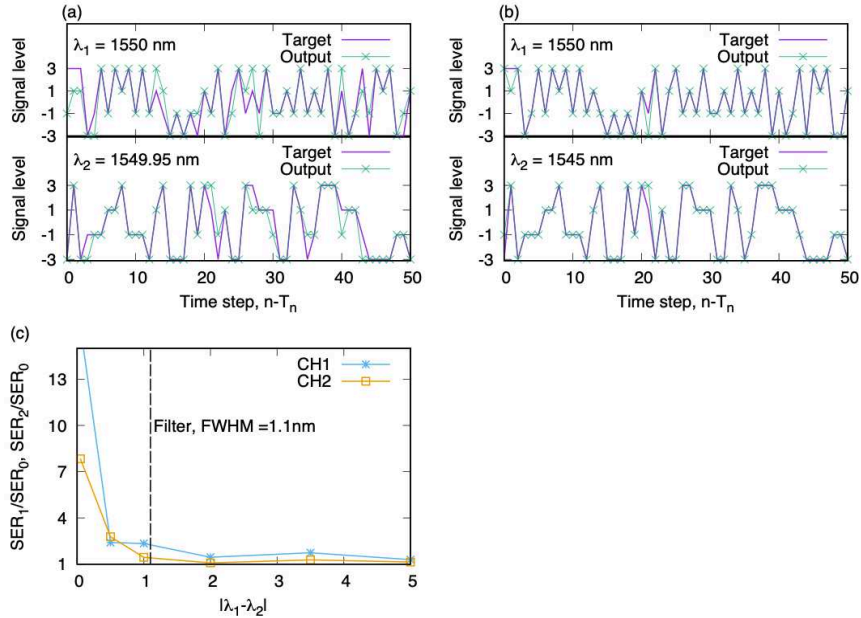}
\caption{\label{fig7}
Results of simultaneous processing for nonlinear channel equalization tasks.
(a) When $\lambda_1-\lambda_2 = 0.05$ nm, large errors occur for each
 channel as a result of the interference of the two signals.
(b) When $\lambda_1-\lambda_2 = 5$ nm, each error rate improves up to the
 level for the case of a single operation.
(c) Symbol error rates [SER$_{1(2)}$ for channel 1(2)] normalized by SER$_0$ as a function of the
 difference in wavelength $|\lambda_1-\lambda_2|$.
Here, SER$_0$ denotes the error rate for the case of a single operation. 
In (a)-(c), the input signals were set to include two white Gaussian
 noises $v_1$ and $v_2$ with zero mean adjusted in power to yield an SNR
 of 30 dB. 
}
\end{figure}

Figure \ref{fig7} shows the parallel processing at a signal input rate
of 12.5 GS/s for this task, where 
the respective readouts for channels 1 and 2 were calculated as  
$\tilde{y}_{1(2)}(n) = \sum_{k=0}^K\sum_i^Nw_{i,k}\phi_i^{\lambda_{1(2)}}[(n-k)\tau]$
with $N = 100$ and $K = 7$ in this experiment.
We used $T_n =$ 3000 samples for training and 3000 samples for testing. 
When the difference between the two wavelengths, $\Delta \lambda = |\lambda_1-\lambda_2|$, 
 is smaller than the bandwidth of the optical filter 
$\Delta \approx 1.1$ nm, the optical signal
 of the wavelength $\lambda_{2(1)}$ cannot be sufficiently eliminated
 with the optical filter 1(2), and their interference causes large
 errors in the parallel computing [Fig. \ref{fig7}(a)]. 
The SERs were worse than that for the single operation, i.e., SER$_0 = 0.022$. 
However, when $\Delta\lambda > \Delta$, the SERs improved, and SER/SER$_0$ $\approx$
1.0 was obtained, i.e., parallel processing can be performed, 
although the original error rate, SER$_0$, was worse than in previous studies.
The low SER$_0$ may be due to the low stability of the multimode fiber
in our experiment; 
it is likely to improve when the multimode fiber is replaced with a
multimode waveguide in a photonic chip.
We consider that further parallel and
multitasking processing will be conducted using optical filters
with a narrower bandwidth.
The number of parallel operations will not be limited in principle. 
This is a unique property of the proposed optical information processing system.
%
%

\section{Summary and discussion}
In this study, we experimentally demonstrated that the use of multidimensional speckle
dynamics in multimode fibers 
enables fast and parallel information processing for time-dependent
signals at input rates of 12.5 GS/s. 
The rate of information processing is limited only by the bandwidths of the phase modulator and 
photodetectors in the present experiment, and can increase further when
those with larger bandwidths are used.   
Larger-scale information processing can in principle be 
achieved through the simultaneous use of the output signals (network nodes)
in both the space and wavelength domains.
In addition, the use of wavelength multiplexing combined with spatial
speckles enables a multitasking operation, as demonstrated in
Sec. \ref{sec5}.  

\begin{figure}[htbp]
\centering\includegraphics[width=12cm]{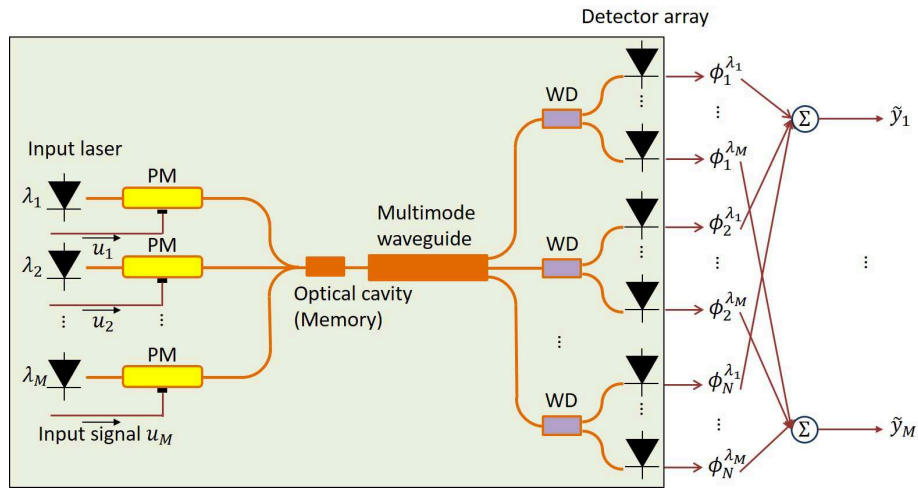}
\caption{\label{fig8}
Conceptual schematic for photonic parallel reservoir computer consisting
 of multiple lasers, phase modulators (PMs), optical cavity for memory,
 wavelength demultiplexers (WDs), and photodetector array. 
}
\end{figure}

Although the present experimental system does not have any memory, the
memory effects can be easily introduced by adding an optical cavity to
store past information, and the system with a memory can be used as a
reservoir computer. 
Figure \ref{fig8} shows an example of the entire photonic architecture for reservoir computing. 
The multimode waveguide with a spiral geometry or interferometer
structure fabricated on a silicon chip can provide a long optical path
length and high NA to generate speckles sensitive to the wavelength
\cite{Piels2017, Redding2016, Paudel2019};
further, it can be used to induce speckle dynamics with a shorter latency in a small footprint. 
Each signal from the multimode waveguide can be split by a splitter
and wavelength demultiplexer and detected by the photodetector array. 
This type of photonic integration will offer a pathway for compact,
parallel, large-scale photonic computers.

\appendix
\section*{Appendix: Speckle-based mapping using a multimode fiber
 \label{appendixA}}
Here, we show a mapping model based on dynamic speckle patterns in a
multimode fiber for a time-dependent signal $u(t)$.
In the experimental setup shown in Fig. \ref{fig1}(a), 
the laser light is phase-modulated using a voltage $V(t)$
proportional to $u(t)$ and sent to a multimode fiber.
The input electric field is given by $\EE_{in}(\rr,t) =
\EE_0(\rr)e^{i(\pi V(t)/V_{\pi}+\omega_0 t)}$, 
where $V_{\pi}$ is the voltage required to produce a $\pi$-phase shift in a phase modulator, 
$\rr = (x,y)$ is the transverse dimension of the propagation, and
$\omega_0$ is the input angular frequency. 
Given that the field can be expressed as the superposition of signals
with angular frequency $\omega$, i.e., 
$\EE_{in}(\rr,t) = \int \hat{\EE}_{in}(\rr,\omega)e^{i\omega t}d\omega$, 
each frequency component after propagating in a multimode fiber 
is written as follows \cite{Redding2013}: 
\begin{align}
\hat{\EE}(\rr,z,\omega) = \sum_m A_m(\omega)\PPsi_{m}(\rr,\omega)
e^{-i\beta_m(\omega)z},
\end{align}
where 
$A_m(\omega)$ is the complex amplitude of the $m$-th guided mode which has the
spatial profile $\PPsi_m(\rr,\omega)$ and propagation constant
$\beta_m(\omega)$ 
for input angular frequency $\omega$, and is determined such that the boundary
condition $\hat{\EE}(\rr,0,\omega) = \hat{\EE}_{in}(\rr,\omega)$ is
satisfied.
Thus, the entire output field at the end of a multimode fiber with length $L$ 
at time $t$ can be expressed by the Fourier transform: 
\begin{align}
    \EE(\rr,L,t) = \int \hat{\EE}(\rr,L,\omega)e^{i\omega t}d\omega,
\end{align}
and the intensity of the field at $\rr_i$ $(i=1,2,\cdots, N)$ is measured 
as $\phi(\rr_i,t)=|\EE(\rr_i,L,t)|^2$. 
As shown in the above model, the mapping from time-dependent signal
$u(t)$ to signal $\phi(\rr_i,t)$ is nonlinear  
and exhibits high dimensionality when $A_m(\omega)\ne 0$ for a
wide range of frequencies $\omega$ and modes $m$, i.e., the input signal
is modulated at fast rates, and many guided modes are excited by the
input field. 

\section*{Funding}
Japan Society for the Promotion of Science (JSPS) (KAKENHI 19H00868,
20H04255, 20K15185);
Japan Science and Technology Agency (JST) (PRESTO JPMJPR19M4);
The Okawa Foundation for Information and Telecommunications;
The Telecommunications Advancement Foundation.

\section*{Disclosures}
The authors declare no conflicts of interest.

\end{document}